\documentclass[conference]{IEEEtran}
\usepackage[letterpaper, left=0.63in, right=0.63in, bottom=1.01in, top=0.75in]{geometry}
\IEEEoverridecommandlockouts
\usepackage[binary-units]{siunitx}
\usepackage[caption=false,font=footnotesize]{subfig}
\usepackage{tabularx}
\usepackage{multirow}
\usepackage{diagbox}
\usepackage{color, colortbl}
\usepackage[binary-units]{siunitx}
\DeclareSIUnit{\bps}{bps}
\usepackage{amsmath,amssymb,amsfonts}
\usepackage{cite}
\usepackage{graphicx}
\usepackage{flushend}

\begin{document}
\title{Intelligent Reflecting Vehicle Surface: A Novel IRS Paradigm for Moving Vehicular Networks}

\author{\IEEEauthorblockN{Wei Jiang\IEEEauthorrefmark{1}\IEEEauthorrefmark{2} and Hans D. Schotten\IEEEauthorrefmark{2}\IEEEauthorrefmark{1}}
\IEEEauthorblockA{\IEEEauthorrefmark{1}Intelligent Networking Research Group, German Research Center for Artificial Intelligence (DFKI)\\Trippstadter Street 122,  Kaiserslautern, 67663 Germany\\
  }
\IEEEauthorblockA{\IEEEauthorrefmark{2}Department of Electrical and Computer Engineering, Technische Universit\"at (TU) Kaiserslautern\\Building 11, Paul-Ehrlich Street, Kaiserslautern, 67663 Germany\\
 }
}



\maketitle

\begin{abstract}
Intelligent reflecting surface (IRS) has recently received much attention from the research community due to its potential to achieve high spectral and power efficiency cost-effectively. In addition to traditional cellular networks, the use of IRS in vehicular networks is also considered. Prior works on IRS-aided vehicle-to-everything communications focus on deploying reflection surfaces on the facades of buildings along the road for sidelink performance enhancement. This paper goes beyond the state of the art by presenting a novel paradigm coined Intelligent Reflecting Vehicle Surface (IRVS). It embeds a massive number of reflection elements on vehicles' surfaces to aid moving vehicular networks in military and emergency communications. We propose an alternative optimization method to optimize jointly active beamforming at the base station and passive reflection across multiple randomly-distributed vehicle surfaces. Performance evaluation in terms of sum spectral efficiency under continuous, discrete, and random phase shifts is conducted. Numerical results reveal that IRVS can substantially improve the capacity of a moving vehicular network.
\end{abstract}

%
\IEEEpeerreviewmaketitle

\section{Introduction}

Intelligent reflecting surface (IRS), also known as reconfigurable intelligent surface, has attracted much attention recently from the research community \cite{Ref_renzo2020smart} due to its potential of realizing high spectral efficiency in a green and cost-efficient manner. IRS is a planar surface consisting of a large number of small, passive, and cheap reflecting elements, each of which is able to independently induce a phase shift to an impinging signal. Unlike conventional wireless techniques that can only \textit{passively} adapt to a radio channel, IRS \textit{proactively} modifies it through controllable reflection, thereby collaboratively achieving smart propagation environment \cite{Ref_wu2019intelligent}. By adaptively adjusting the phase shifts of IRS, the reflected signals can be added either constructively with the direct signal to increase signal strength, or destructively to suppress co-channel interference. Since its reflecting elements (e.g., low-cost printed dipoles or positive-intrinsic-negative (PIN) diodes) only passively reflect the impinging signals, radio-frequency (RF) chains for signal transmission and reception become unnecessary \cite{Ref_zhi2021uplink}. Thus, it can be implemented with orders-of-magnitude lower hardware cost and power consumption than active antenna arrays. Consequently, it is widely recognized that IRS is able to serve as a key technological enabler for the upcoming sixth-generation (6G) wireless system to meet more stringent performance requirements than its predecessor \cite{Ref_jiang2021road}.

In addition to traditional cellular networks, the use of IRS in vehicular networks is also considered. Prior works on IRS-aided vehicle-to-everything communications \cite{Ref_chen2022reconfigurable} focus on deploying reflection surfaces on the facades of buildings along the road for sidelink performance enhancement. For example, \cite{Ref_alhilo2022reconfigurable} leverages IRS to enable indirect transmission for dark zones of roadside units. \cite{Ref_chen2020resource} studies the resource allocation for IRS to maximize the capacity of vehicle-to-infrastructure link while guaranteeing the minimum quality-of-service of vehicle-to-vehicle links.   Reflecting elements are generally low-profile, lightweight, and conformal geometric. Hence, a surface can be practically fabricated in arbitrary shapes to cater for a wide variety of deployment scenarios and be integrated into wireless networks transparently as auxiliary equipment. Taking advantage of this technical feature, this paper goes beyond the state of the art by presenting a novel paradigm coined Intelligent Reflecting Vehicle Surface (IRVS). It embeds a massive number of reflection elements on vehicles' surface to aid moving vehicular networks in military and emergency communications. We propose an alternative optimization method to optimize jointly active beamforming at the base station (BS) and passive reflection over vehicle surfaces. Performance evaluation in terms of sum spectral efficiency under continuous, discrete, and random phase shifts is conducted through simulations. Numerical results reveal that the introduction of IRVS can substantially improve the capacity of a moving vehicular network.

The remainder of this paper is organized as follows: Section II introduces the system model of IRVS-aided moving vehicular networks. Section III presents the joint optimization design to maximize sum spectral efficiency. Simulation setup and some examples of numerical results are demonstrated in Section IV. Finally, the conclusions are drawn in Section V.

\section{System Model}
We consider a moving vehicular network where a multi-antenna BS serves a number of users with the aid of multiple surfaces. The BS is mounted on the top of a vehicle that is dedicated to providing communications service to its nearby humans or vehicles, while reflecting elements are installed on the surfaces of some other vehicles.  This scenario corresponds to a platoon of interconnected combat vehicles and battle staff in the battlefield, or a group of special vehicles and rescue team members for disaster relief and emergency events. Unlike cellular networks, where intelligent surfaces are deliberately deployed on the facades of buildings, walls, and ceilings in a stationary manner, this paper is interested in a moving vehicular network where a massive number of reflection elements are embedded on vehicles' surfaces. Since reflection elements are small, lightweight, and passive, it is easy to fabricate surfaces in any geometry, size, orientation, and arrangement. There are a plenty of large vehicles involved in either the military or emergency applications, so the implementation of IRVS is feasible.

\begin{figure}[!t]
    \centering
    \includegraphics[width=0.46\textwidth]{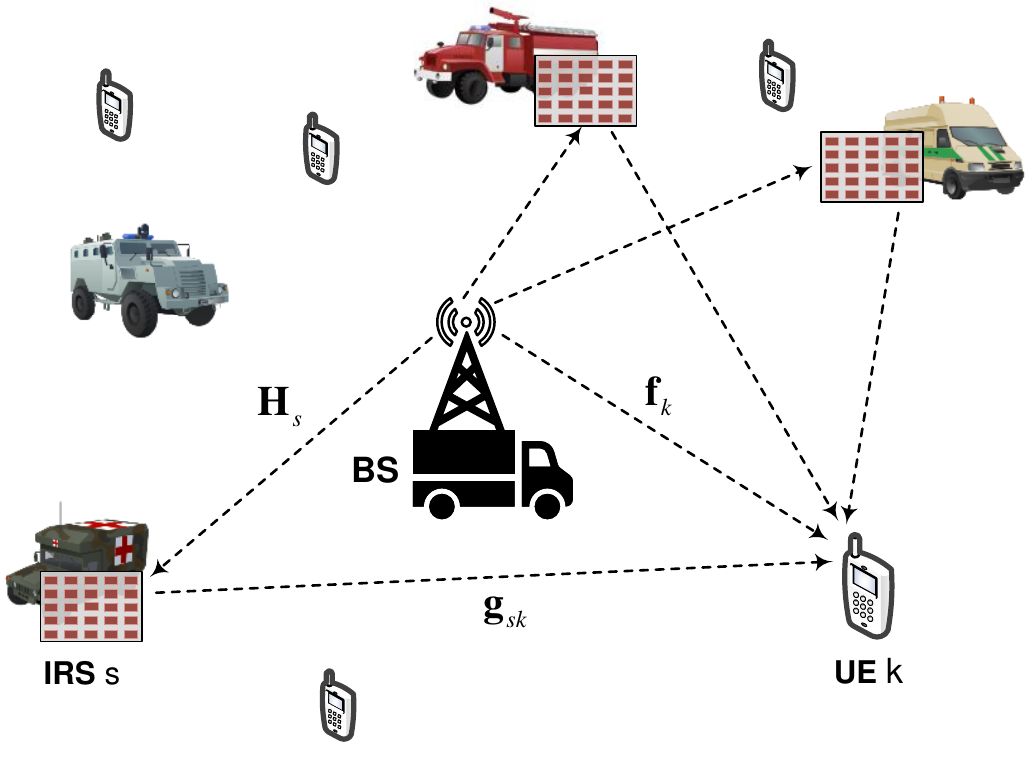}
    \caption{Schematic diagram of a moving vehicular network, consisting of a multi-antenna BS mounted on a vehicle, $K$ single-antenna UE for humans or vehicles, and $S$ reconfigurable surfaces embedded on vehicles.  }
    \label{diagram:system}
\end{figure}

\figurename \ref{diagram:system} illustrates the schematic diagram of an IRVS-assisted vehicular network, where $S$ vehicle surfaces are deployed to assist the transmission from an $N_b$-antenna BS to $K$ single-antenna user equipment (UE). Due to the orientation of vehicles, some elements might not be able to reflect the impinging signals. Without losing generality, we assume the $s^{th}$ surface has $N_s$ effective reflecting elements, where $s\in \mathcal{S} \triangleq\{1,2,\ldots,S\}$.
Since the IRVS is a passive device, time-division duplexing (TDD) operation is usually adopted to simplify channel estimation. The users send pilot signals in the training period, such that the BS can estimate the uplink channel state information (CSI), which is used for optimizing downlink data transmission due to channel reciprocity.
To characterize the theoretical performance, the analysis hereinafter is conducted under the assumption that the CSI of all involved channels is perfectly known at the BS, as most prior works \cite{Ref_wu2019intelligent, Ref_renzo2020reconfigurable, Ref_renzo2020smart, Ref_wang2020channel,Ref_zhi2021uplink, Ref_jiang2022dualbeam}. In addition, we assume narrowband communications, where the channels follow frequency-flat block fading. A wideband channel suffering from frequency selectivity can be transformed into a set of narrowband channels through OFDM \cite{Ref_jiang2016ofdm}, making the assumption of flat fading reasonable.

Since the direct paths from either the BS or the IRVS to UEs may be blocked, the corresponding small-scale fading follows Rayleigh distribution. Consequently, the channel gain between antenna element $n_b\in  \{1,2,\ldots,N_b\}$ and user $k\in \mathcal{K} \triangleq\{1,2,\ldots,K\}$ is a circularly symmetric complex Gaussian random variable with zero mean and variance $\sigma_k^2$, i.e., $f_{kn_b} \sim \mathcal{CN}(0,\sigma_k^2)$. Thus, the channel vector from the BS to the $k^{th}$ UE is denoted by
\begin{align}
    \mathbf{f}_{k}=\Bigl[f_{k1},f_{k2},\ldots,f_{kN_b}\Bigr]^T.
\end{align}
Similarly, the channel between the $s^{th}$ IRVS to the $k^{th}$ UE is modeled as an $N_s\times 1$ vector:
\begin{align}
    \mathbf{g}_{sk}=\Bigl[g_{sk,1},g_{sk,2},\ldots,g_{sk,N_s}\Bigr]^T,
\end{align}
where $g_{sk,n}\sim \mathcal{CN}(0,\sigma_{sk}^2)$ is the channel gain between element $n\in \mathcal{N}_s \triangleq \{1,2,\ldots,N_s\}$ of the $s^{th}$ surface and user $k$, and $\sigma_{sk}^2$ is the large-scale fading between surface $s$ and user $k$.
We write
\begin{equation}
    \mathbf{h}_{sn}=\Bigl[h_{sn,1},h_{sn,2},\ldots,h_{sn,N_b}\Bigr]^T
\end{equation} to denote the channel vector between the BS and the $n^{th}$ reflecting element of surface $s$. So the channel matrix from the BS to the $s^{th}$ IRVS is expressed as $\mathbf{H}_s\in \mathbb{C}^{N_s\times N_b}$, where the $n^{th}$ row of $\mathbf{H}_s$ equals to $\mathbf{h}_{sn}^T$.  In contrast to hand-held UE, a surface mounted on a vehicle is generally high enough to get the line-of-sight (LOS) path of the BS without any blockage, resulting in Rician fading, i.e.,
\begin{equation}\label{EQNIRQ_LSFadingdirect}
    \mathbf{H}_s=\sqrt{\frac{\Gamma\sigma_s^2}{\Gamma+1}}\mathbf{H}_{LOS} + \sqrt{\frac{\sigma_s^2}{\Gamma+1}}\mathbf{H}_{NLOS},
\end{equation}
with the Rician factor $\Gamma$, the LOS component $\mathbf{H}_{LOS}$,  the multipath component $\mathbf{H}_{NLOS}$ consisting of independent entries that follow $\mathcal{CN}(0,1)$, and the BS-IRVS path loss $\sigma_s^2$, $s\in \mathcal{S}$.

Each reflecting surface is equipped with a smart controller, which can adaptively adjust the phase shift of each element in terms of the CSI acquired through periodic channel estimation \cite{Ref_wang2020channel}. We write $c_{sn}=\beta_{sn} e^{j\phi_{sn}}$ to denote the reflection coefficient of the $n^{th}$ element of the $s^{th}$ surface, with the induced phase shift $\phi_{sn}\in [0,2\pi)$ and the amplitude attenuation $\beta_{sn}\in [0,1]$. As revealed by \cite{Ref_wu2019intelligent}, the optimal value of reflecting attenuation is $\beta_{sn}=1$, $\forall s, n $  to maximize the received power and simplify the hardware implementation.  Because of severe path loss, the signals that are reflected by the IRVS twice or more are negligible. By ignoring impairments such as channel aging \cite{Ref_jiang2021impactcellfree} and phase noise \cite{Ref_jiang2022impact}, the $k^{th}$ UE observes the received signal
\begin{equation} \label{eqn_systemModel}
    r_k=\sqrt{P_d}\Biggl(\sum_{s=1}^{S} \sum_{n=1}^{N_s} g_{sk,n} e^{j\phi_{sn}} \mathbf{h}_{sn}^T + \mathbf{f}_{k}^T \Biggr) \mathbf{x} + n_k,
\end{equation}
where $\mathbf{x}$ denotes the vector of transmitted signals over the BS antenna array, $P_d$ expresses the power constraint of the BS, $n_k$ is additive white Gaussian noise (AWGN) with zero mean and variance $\sigma_n^2$, namely $n_k\sim \mathcal{CN}(0,\sigma_n^2)$. Define the diagonal phase-shift matrix for surface $s$ as
\begin{equation}
    \boldsymbol{\Phi}_s=\mathrm{diag}\Bigl\{e^{j\phi_{s1}},e^{j\phi_{s2}},\ldots,e^{j\phi_{sN_s}}\Bigr\}, \:\:\forall s\in \mathcal{S},
\end{equation} \eqref{eqn_systemModel} can be rewritten in matrix form as
\begin{equation} \label{EQN_IRS_RxSignal_Matrix}
    r_k= \sqrt{P_d}\left(\sum_{s=1}^{S} \mathbf{g}_{sk}^T \boldsymbol{\Phi}_s \mathbf{H}_s +\mathbf{f}_k^T\right)\mathbf{x} +n_k.
\end{equation}

\section{Joint Optimization Design}
Due to the lack of frequency-selective passive beamforming, namely the set of phase shifts cannot be tuned differently over different frequency subchannels, the use of IRVS imposes a fundamental challenge on frequency-division multiple access (FDMA). In \cite{Ref_zheng2020intelligent_COML}, the authors revealed that time-division multiple access (TDMA) is superior to FDMA with a substantial spectral-efficiency gain. Therefore, we adopt TDMA as the multiple access scheme to analyze the IRVS-aided vehicular communications, which orthogonally divides a radio frame into $K$ time slots. Each user transmits over the entire bandwidth but cyclically accesses its assigned slot. At the header of each frame, the BS processes the uplink pilot signals to estimate the CSI. The duration of a radio frame is usually set to less than the channel coherence time, so that the CSI keeps constant for the whole frame.

At the $k^{th}$ slot, the BS applies linear beamforming $\mathbf{w}_k\in \mathbb{C}^{N_b\times 1}$, where $\|\mathbf{w}_k\|^2\leqslant 1$, to send the information-bearing symbol $x_k$ with zero mean and unit variance, i.e., $\mathbb{E}\left[|x_k|^2\right]=1$, intended for a general user $k$. According to \cite{Ref_zhang2018space}, the switching frequency of reflection elements made by PIN diodes can reach 5 megahertz (\si{\mega\hertz}), corresponding to the switching time of \SI{0.2}{\micro\second}, much smaller than the typical  coherence time on the order of \si{\milli\second}. It implies that the set of phase shifts can be adjusted specifically for the active user at each time slot. We write $\phi_{sn}[k]\in [0,2\pi)$ to denote the phase shift of the $n^{th}$ element of surface $s$ at slot $k$, and define the \textit{time-selective} phase-shift matrix as
\begin{equation}
    \boldsymbol{\Phi}_s[k]=\mathrm{diag}\Bigl\{e^{j\phi_{s1}[k]},\ldots,e^{j\phi_{sN_s}[k]}\Bigr\},\:\:k \in \mathcal{K}, s\in \mathcal{S}.
\end{equation}
Substituting $\mathbf{x}=\mathbf{w}_k x_k$ and $\boldsymbol{\Phi}_s=\boldsymbol{\Phi}_s[k]$ into \eqref{EQN_IRS_RxSignal_Matrix}, we obtain the observation of user $k$ at slot $k$ as
\begin{equation}
    r_k= \sqrt{P_d}\left(\sum_{s=1}^{S} \mathbf{g}_{sk}^T \boldsymbol{\Phi}_s[k] \mathbf{H}_s +\mathbf{f}_k^T\right)\mathbf{w}_k x_k +n_k.
\end{equation}

\begin{figure}[!t]
    \centering
    \includegraphics[width=0.46\textwidth]{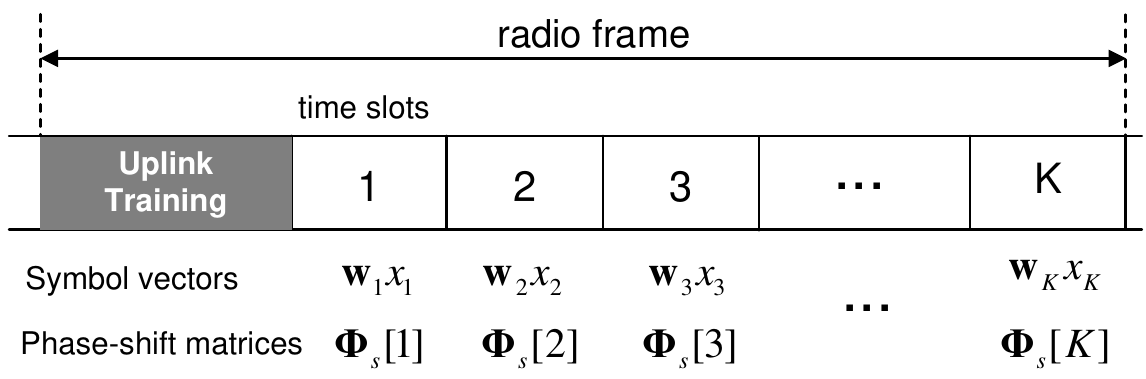}
    \caption{The structure of a TDMA frame.  }
    \label{diagram:TDMA}
\end{figure}

By jointly optimizing $\mathbf{w}_k$ and $\boldsymbol{\Phi}_s[k]$, the instantaneous signal-to-noise ratio (SNR) of user $k$, i.e.,
\begin{equation} \label{IRS_EQN_spectralEfficiency}
    \gamma_k=\frac{P_d \left|\Bigl(\sum_{s=1}^S \mathbf{g}_{sk}^T \boldsymbol{\Phi}_s[k] \mathbf{H}_s +\mathbf{f}_k^T\Bigr)\mathbf{w}_k\right|^2 }{\sigma_n^2}
\end{equation}
can be maximized, formulating the following optimization
\begin{equation}
\begin{aligned} \label{eqnIRS:optimizationMRTvector}
\max_{\boldsymbol{\Phi}_s[k],\:\mathbf{w}_k}\quad &  \left|\left(\sum_{s=1}^S \mathbf{g}_{sk}^T \boldsymbol{\Phi}_s[k] \mathbf{H}_s +\mathbf{f}_k^T\right)\mathbf{w}_k\right|^2 \\
\textrm{s.t.} \quad & \|\mathbf{w}_k\|^2\leqslant 1\\
  \quad & \phi_{sn}[k]\in [0,2\pi), \: \forall s\in \mathcal{S}, \forall n\in\mathcal{N}_s, \forall k\in \mathcal{K}.
\end{aligned}
\end{equation}
This formula is non-convex because the objective function is not jointly concave with respect to $\boldsymbol{\Phi}_s[k]$ and $\mathbf{w}_k$. To solve this problem, we can apply alternating optimization that alternately optimizes $\boldsymbol{\Phi}_s[k]$ and $\mathbf{w}_k$ in an iterative manner \cite{Ref_wu2019intelligent}.
Given an initialized transmit vector $\mathbf{w}_k^{(0)}$, \eqref{eqnIRS:optimizationMRTvector} is simplified to
\begin{equation}  \label{eqnIRS:optimAO}
\begin{aligned} \max_{\boldsymbol{\Phi}_s[k]}\quad &  \left|\left(\sum_{s=1}^S\mathbf{g}_{sk}^T \boldsymbol{\Phi}_s[k] \mathbf{H}_s +\mathbf{f}_{k}^T\right)\mathbf{w}_k^{(0)}\right|^2\\
\textrm{s.t.}  \quad & \phi_{sn}[k]\in [0,2\pi), \: \forall s\in \mathcal{S}, \forall n\in\mathcal{N}_s, \forall k\in \mathcal{K}.
\end{aligned}
\end{equation}
The objective function is still non-convex but it enables a closed-form solution through applying the well-known triangle inequality
\begin{align} \nonumber
    &\left|\left(\sum_{s=1}^S\mathbf{g}_{sk}^T \boldsymbol{\Phi}_s[k] \mathbf{H}_s +\mathbf{f}_{k}^T\right)\mathbf{w}_k^{(0)}\right|\\ &\leqslant \left|\sum_{s=1}^S\mathbf{g}_{sk}^T \boldsymbol{\Phi}_s[k] \mathbf{H}_s \mathbf{w}_k^{(0)}\right| + \left|\mathbf{f}_{k}^T\mathbf{w}_k^{(0)}\right|.
\end{align}
The equality achieves if and only if
\begin{equation} \label{EQN_IRS_angleFormular}
    \arg\left (\sum_{s=1}^S\mathbf{g}_{sk}^T \boldsymbol{\Phi}_s[k] \mathbf{H}_s \mathbf{w}_k^{(0)}\right)= \arg\left(\mathbf{f}_{k}^T\mathbf{w}_k^{(0)}\right)\triangleq \varphi_{0k},
\end{equation}
where $\arg(\cdot)$ stands for the phase of a complex scalar.

Define an $N_r\times 1$ vector to model the overall channel from all surfaces to the $k^{th}$ UE
\begin{equation}
    \mathbf{g}_k=\left[\mathbf{g}_{1k}^T,\mathbf{g}_{2k}^T,\ldots,\mathbf{g}_{Sk}^T \right]^T,
\end{equation}
where $N_r=\sum_{s=1}^S N_s$ is the total number of reflecting elements over all surfaces,  and the overall phase-shifts matrix at slot $k$
\begin{equation}
    \boldsymbol{\Phi}[k]=\mathrm{diag}\biggl\{e^{j\phi_{11}[k]},e^{j\phi_{12}[k]},\ldots,e^{j\phi_{SN_S}[k]}\biggr\},
\end{equation}
and an $N_r\times N_b$ channel matrix from the BS to all surfaces as
\begin{equation} \mathbf{H}=\Bigl[
\mathbf{H}_1^T, \mathbf{H}_2^T, \cdots, \mathbf{H}_S^T
\Bigr]^T. \end{equation}
Thus, \eqref{EQN_IRS_angleFormular} is transformed to
\begin{equation}
    \arg\left (\mathbf{g}_{k}^T \boldsymbol{\Phi}[k] \mathbf{H} \mathbf{w}_k^{(0)}\right)= \arg\left(\mathbf{f}_{k}^T\mathbf{w}_k^{(0)}\right)\triangleq \varphi_{0k}.
\end{equation}

Defining $\mathbf{q}_k=\Bigl[q_{11}[k],q_{12}[k],\ldots,q_{SN_S}[k]\Bigr]^H$ with
\begin{equation}
    q_{sn}[k]=e^{j\phi_{sn}[k]}
\end{equation}
and
\begin{equation}
    \boldsymbol{\chi}_k=\mathrm{diag}(\mathbf{g}_k^T)\mathbf{H}\mathbf{w}_k^{(0)},
\end{equation}
we have
\begin{equation}
    \mathbf{g}_{k}^T \boldsymbol{\Phi}[k] \mathbf{H} \mathbf{w}_k^{(0)}= \mathbf{q}_k^H \boldsymbol{\chi}_k.
\end{equation}
Ignoring the constant term $\bigl|\mathbf{f}_{k}^T\mathbf{w}_k^{(0)}\bigr|$, \eqref{eqnIRS:optimAO} is transformed to
\begin{equation}  \label{eqnIRS:optimizationQ}
\begin{aligned} \max_{\boldsymbol{\mathbf{q}_k}}\quad &  \Bigl|\mathbf{q}_k^H\boldsymbol{\chi}_k\Bigl|\\
\textrm{s.t.}  \quad & |q_{sn}[k]|=1, \: \forall s\in \mathcal{S}, n\in \mathcal{N}_s, k\in \mathcal{K},\\
  \quad & \arg(\mathbf{q}_k^H\boldsymbol{\chi}_k)=\varphi_{0k}.
\end{aligned}
\end{equation}
The surfaces should be tuned such that the reflected signals and the signals over the direct link are phase-aligned to achieve coherent combining. Consequently, the solution for \eqref{eqnIRS:optimizationQ} is
\begin{equation} \label{eqnIRScomplexityQ}
    \mathbf{q}^{(1)}_k=e^{j\left(\varphi_{0k}-\arg(\boldsymbol{\chi}_k)\right)}=e^{j\left(\varphi_{0k}-\arg\left( \mathrm{diag}(\mathbf{g}_k^T)\mathbf{H}\mathbf{w}_k^{(0)}\right)\right)}.
\end{equation}
Once the required phases, i.e.,
\begin{equation}
    \boldsymbol{\Phi}^{(1)}[k]=\mathrm{diag}\left\{\mathbf{q}^{(1)}_k\right\},
\end{equation}
are determined, the optimization is alternated to update $\mathbf{w}_k$. The BS can apply maximal-ratio transmission (MRT) a.k.a. matched filtering to maximize the strength of a desired signal, resulting in
\begin{equation}  \label{EQN_IRS_TXBF}
    \mathbf{w}_k^{(1)} = \frac{\Bigl(\mathbf{g}_{k}^T \boldsymbol{\Phi}^{(1)}[k] \mathbf{H} +\mathbf{f}_{k}^T\Bigr)^H}{\Bigl\|\mathbf{g}_{k}^T \boldsymbol{\Phi}^{(1)}[k] \mathbf{H} +\mathbf{f}_{k}^T\Bigr\|}.
\end{equation}

After the completion of the first optimization iteration, the BS gets $\boldsymbol{\Phi}^{(1)}[k]$ and $\mathbf{w}_k^{(1)}$, which serve as the initial input for the second iteration to derive $\boldsymbol{\Phi}^{(2)}[k]$ and $\mathbf{w}_k^{(2)}$.
This process iterates until the convergence is achieved with the optimal transmit vectors and phase-shift matrices denoted by $\mathbf{w}_k^{\star}$ and  $\boldsymbol{\Phi}^{\star}[k]$, $k\in \mathcal{K}$, respectively. Then, we can derive the achievable spectral efficiency of user $k$ as
\begin{align} \label{IRS_EQN_TDMA_SE} \nonumber
    C_k&=\frac{1}{K}\log\left(1+\frac{P_d \Bigl|\bigl(\mathbf{g}_{k}^T \boldsymbol{\Phi}^\star[k] \mathbf{H} +\mathbf{f}_k^T\bigr)\mathbf{w}_k^\star \Bigr|^2 }{\sigma_n^2} \right)\\
    &=\frac{1}{K}\log\left(1+\biggl \| \mathbf{g}_{k}^T \boldsymbol{\Phi}^\star[k] \mathbf{H} +\mathbf{f}_k^T \biggr\|^2 \frac{P_d}{\sigma_n^2} \right),
\end{align}
and the sum rate of this IRVS-aided vehicular system as
\begin{align} \nonumber
    C&=\frac{1}{K} \sum_{k=1}^K C_k\\
    &=\frac{1}{K} \sum_{k=1}^K \log\left(1+\biggl\| \mathbf{g}_{k}^T \boldsymbol{\Phi}^\star[k] \mathbf{H} +\mathbf{f}_k^T \biggr \|^2 \frac{P_d}{\sigma_n^2} \right).
\end{align}

\subsection{Discrete Phase Shifts}
Although continuous phase shifts are beneficial for optimizing performance, it is practically difficult to implement. PIN diodes have been widely adopted for fabricating reflection elements due to fast response time, small reflection loss, and relatively low hardware and energy costs. By adding different biasing voltages, a PIN diode is switched to either ON or OFF state, inducing a phase-shift difference of $\pi$. Finer controlling is realized by integrating multiple diodes. For example, implementing $8$ discrete phase shifts needs at least $\log_28=3$ PIN diodes \cite{Ref_wu2021intelligent}.
Higher resolution imposes not only high hardware cost and design complexity but also a large burden of wireless or wired backhaul between the BS and smart controllers. Since a surface usually contains a large number of reflecting elements, it is practical to implement only a finite number of discrete phase shifts denoted by $\theta_{sn}$, $\forall s\in \mathcal{S}, n\in \mathcal{N}_s$. Let $b$ denote the number of bits for representing $L=2^b$ different levels,
the set of discrete phase shifts can be expressed as
\begin{equation}
    \mathcal{D} = \Bigl\{0, \triangle \theta, 2\triangle \theta, \ldots, (L-1) \triangle \theta \Bigr\}
\end{equation}
with $\triangle \theta=2\pi/L$. To optimize active beamforming at the BS and discrete reflection at the IRVS, we propose to apply a two-step alternative optimization method, which gets the optimal continuous phase shifts $\boldsymbol{\Phi}^{\star}[k]$ first, and then quantizes each phase shift to its nearest discrete value. Each phase shift is quantized into a finite number of phase-control bits before sending to its corresponding smart controller through a wireless or wired link. The quantization can be simply implemented through a mid-tread uniform quantizer as
\begin{equation}
    \theta_{sn}^\star = \triangle \theta \left \lfloor  \frac{\phi_{sn}^\star}{\triangle \theta } + 0.5\right\rfloor,
\end{equation}
where the notation $\lfloor \cdot \rfloor$  denotes the floor function.

\subsection{Random Phase Shifts}
In addition to the joint optimization of passive reflection and active beamforming, we study the simplest solution as a baseline. Randomly setting the phase shifts of all reflecting elements denoted by $\boldsymbol{\Phi}_r[k]$, $k\in \mathcal{K}$, each entry of which takes value uniformly and randomly from continuous phase shifts $[0,2\pi)$, or $\boldsymbol{\Theta}_r[k]$, $k\in \mathcal{K}$ that takes value uniformly and randomly from discrete phase shifts $\mathcal{D}$. Then, its transmit beamforming is optimized as
\begin{equation}
    \mathbf{w}_{k}^{\star} = \frac{\Bigl(\mathbf{g}_{k}^T \boldsymbol{\Theta}_{r}[k] \mathbf{H} +\mathbf{f}_{k}^T\Bigr)^H}{\Bigl\|\mathbf{g}_{k}^T \boldsymbol{\Theta}_{r}[k] \mathbf{H} +\mathbf{f}_{k}^T\Bigr\|}.
\end{equation}
Random reflection does not need CSI acquisition, reflection optimization, and backhauling, which not only simplify the system design but also improve the network robustness, with the price of some performance loss.

\begin{figure}[!ht]
    \centering
    \includegraphics[width=0.46\textwidth]{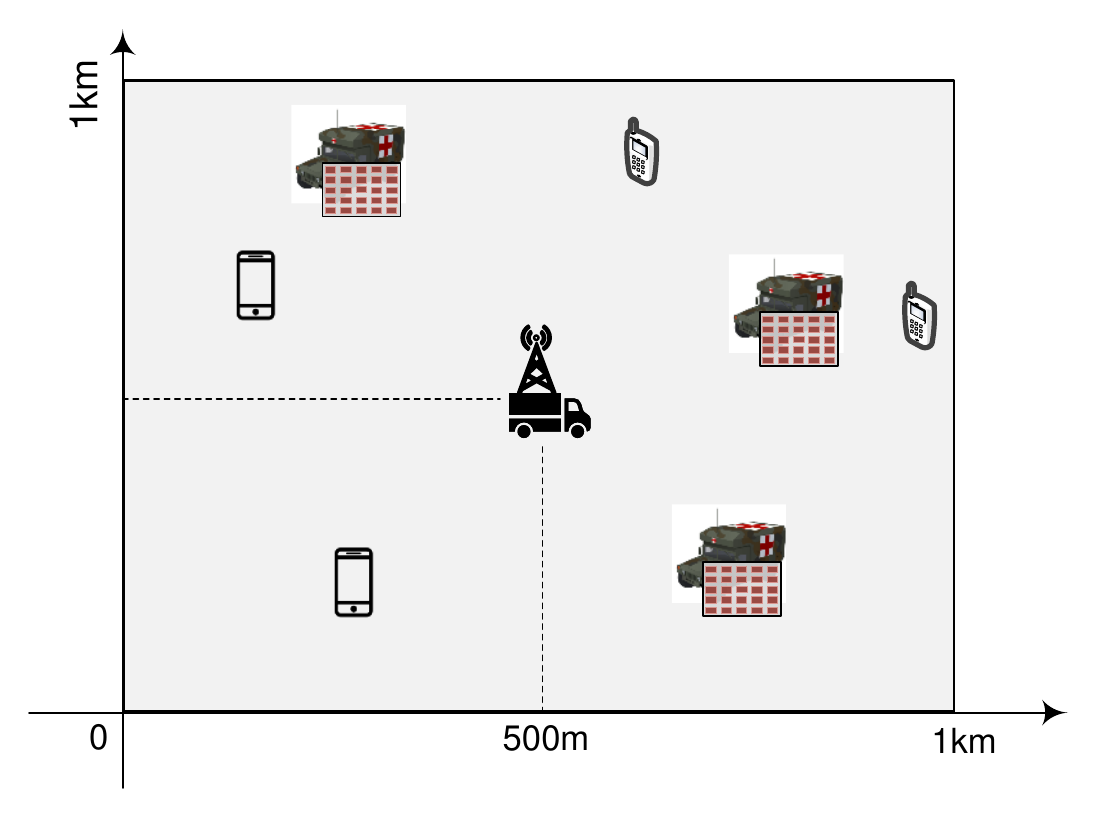}
    \caption{Simulation scenario of an IRVS-aided vehicular network, where the BS serves $K$ users in an area of $\SI{1}{\kilo\meter}\times \SI{1}{\kilo\meter}$ with the aid of $S$ surfaces. Both the users and vehicle surfaces are randomly distributed within the area.}
    \label{diagram:Simulation}
\end{figure}

\section{Performance Evaluation}
\begin{figure*}[!t]
\centerline{
\subfloat[]{
\includegraphics[width=0.45\textwidth]{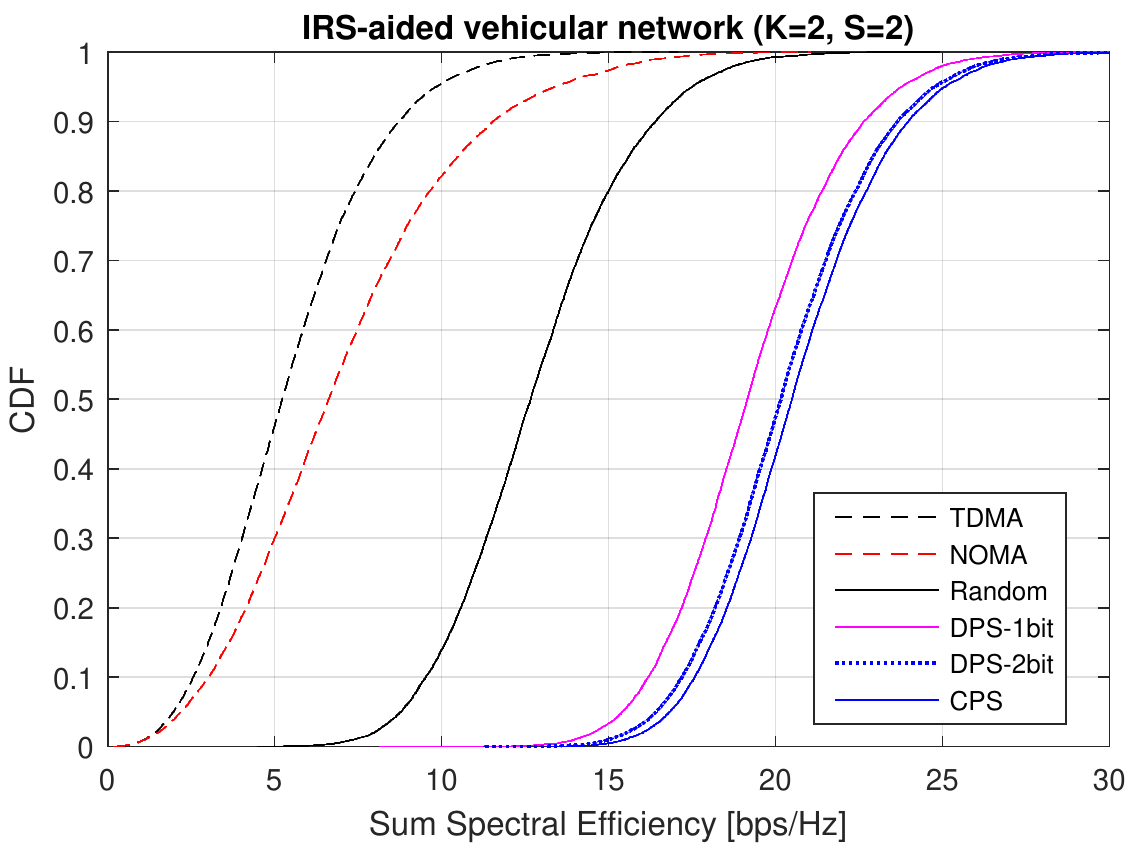}
\label{Fig_results1}
}
\hspace{10mm}
\subfloat[]{
\includegraphics[width=0.45\textwidth]{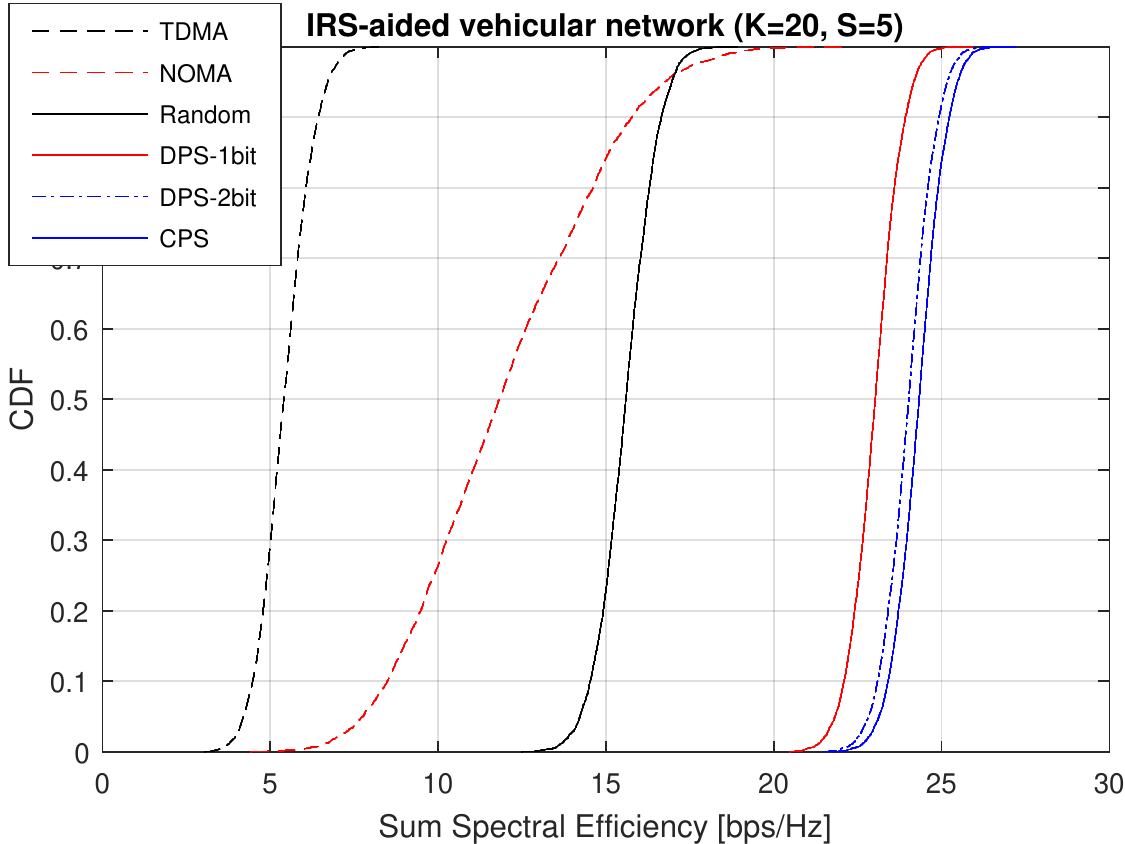}
\label{Fig_results2}
}}
\caption{Simulation results of an IRVS-aided vehicular network under continuous (\textit{CPS}), discrete (\textit{DPS}), and random phase shifts (\textit{RPS}): (a) CDFs of the sum spectral efficiency with two users and two surfaces, and (b) CDFs of the sum spectral efficiency with twenty users and five surfaces. }
\label{Fig_Result}
\end{figure*}

\begin{table*}[!t]
\renewcommand{\arraystretch}{1.3}
\scriptsize
\caption{$95\%$-likely and $50\%$-likely sum spectral efficiency.}
\label{table_IRSdatarates}
\centering
\begin{tabular}{|c|l|l|l|l|l|l|l|l|l|l|l|l|}
\hline
\multirow{2}{*}{\backslashbox{Sum Rate}{}} &\multicolumn{6}{|c|}{$K=2$,\:$S=2$}&\multicolumn{6}{|c|}{$K=20$,\:$S=5$}\\ \cline{2-13}
&TDMA& NOMA & RPS & DPS-1bit &DPS-2bit&CPS& TDMA& NOMA & RPS & DPS-1bit &DPS-2bit&CPS\\ \hline
$95\%$-likely [bps/Hz] &1.98&2.21&8.78&15.42&16.43&16.82&4.20&7.76  &14.23&21.83&22.83&23.14 \\ \hline
Median [bps/Hz] &5.23&6.62&12.67&19.15&20.17&20.49&5.40& 11.82 &15.60&23.03&24.02&24.33\\ \hline
\end{tabular}
\end{table*}
This section first presents the simulation setup and then illustrates some representative numerical results of the IRVS-aided vehicular system in terms of achievable sum spectral efficiency.
As shown in \figurename \ref{diagram:Simulation}, we consider a geographical area with the dimension of $\SI{1}{\kilo\meter}\times \SI{1}{\kilo\meter}$ covered by a moving vehicular system. The BS equipped with $N_b=16$ antennas is located at the central point $(500,500)$ of the coordinate system, while $S=2$ or $5$ vehicle surfaces and $K=2$ or $20$ users are randomly, uniformly distributed in this area. Without losing generality, each surface is assumed to contain $N_s=200$, $\forall s\in \mathcal{S}$ reflecting elements.  Referring to the practical LTE and NR specifications, the maximum transmit power of BS is selected to be $P_d=20\mathrm{W}$ over a signal bandwidth of $B_w=20\mathrm{MHz}$, corresponding to a power density of $-30\mathrm{dBm}$.   The variance of white noise  is figured out by $\sigma_n^2=\kappa\cdot B_w\cdot T_0\cdot N_f$ with the Boltzmann constant $\kappa$, temperature $T_0=290 \mathrm{Kelvin}$, and the noise figure  $N_f=9\mathrm{dB}$.

The distance-dependent large-scale fading of the BS-UE and IRVS-UE channels, i.e., the variances $\sigma_k^2$ and $\sigma_{sk}^2$, can be computed by
$10^\frac{\mathcal{P}+\mathcal{S}}{10}$ with path loss $\mathcal{P}$ and shadowing fading $\mathcal{S}\sim \mathcal{N}(0,\sigma_{sd}^2)$. As  \cite{Ref_jiang2021cellfree}, the path loss can be calculated by the COST-Hata model, i.e.,
\begin{equation} \label{eqn:CostHataModel}
    \mathcal{P}=
\begin{cases}
-L-35\log_{10}(d), &  d>d_1 \\
-L-15\log_{10}(d_1)-20\log_{10}(d), &  d_0<d\leq d_1 \\
-L-15\log_{10}(d_1)-20\log_{10}(d_0), &  d\leq d_0
\end{cases},
\end{equation}
where $d$ represents the propagation distance,  $d_0$ and $d_1$ are the break points of the three-slope model, and  \begin{IEEEeqnarray}{ll}
 L=46.3&+33.9\log_{10}\left(f_c\right)-13.82\log_{10}\left(h_{S}\right)\\ \nonumber
 &-\Bigl[1.1\log_{10}(f_c)-0.7\Bigr]h_{T}+1.56\log_{10}\left(f_c\right)-0.8,
\end{IEEEeqnarray} where the carrier frequency $f_c=1.9\mathrm{GHz}$, the antenna height of BS $h_{S}=5\mathrm{m}$, the height of IRVS $h_S=\SI{3}{\meter}$, and the antenna height of UE $h_{T}=1.65\mathrm{m}$.
The break points of the three-slope model in (\ref{eqn:CostHataModel}) take values $d_0=10\mathrm{m}$ and $d_1=50\mathrm{m}$, while the standard derivation for shadowing fading is $\sigma_{sd}=8\mathrm{dB}$. In contrast, the path loss for the LOS channel between the BS and IRVS can be computed through
\begin{equation}
    \sigma_s^2=\frac{L_0}{d^{-\alpha}},
\end{equation}
where $L_0=\SI{-30}{\decibel}$ is the path loss at the reference distance of \SI{1}{\meter}, $\alpha=2.5$ means the path-loss exponent, and the Rician factor in \eqref{EQNIRQ_LSFadingdirect} is set to $\Gamma=5$.

Cumulative distribution functions (CDFs) of sum spectral efficiency in bits-per-second per hertz (\si{\bps\per\hertz^{}}) are evaluated as the performance metric for the IRVS-aided vehicular system. Simulation results when the phase shifts are \textit{continuous},  \textit{discrete} with $b=1$ and $b=2$ phase-control bits, and \textit{random} are provided.
To highlight the performance gain of employing vehicle surfaces, two schemes are applied as the baseline for performance comparison:
\begin{itemize}
    \item Conventional TDMA without the aid of IRVS, where the BS sends the information-bearing symbol $x_k$ towards the $k^{th}$ user at the $k^{th}$ slot. The antenna array applies MRT by setting $\mathbf{w}_k^\star=\mathbf{f}_k^*/\|\mathbf{f}_k\|$ to achieve matched filtering in the BS-UE direct link.
    \item Non-orthogonal multiple access (NOMA): at the transmitter side, all information symbols are superimposed into a single waveform $\mathbf{x}=\sum_{k=1}^K \sqrt{\alpha_k } \mathbf{w}_{k} x_k$,
where  $\alpha_k$ represents the power allocation coefficient. We set $\alpha_k=2k/K(K+1)$,  $\forall k\in \mathcal{K}$ subjecting to $\sum_{k=1}^K\alpha_k= 1$. The optimal order of successive interference cancellation (SIC) is detecting the user with the weakest channel gain to the user with the strongest channel gain. Each user decodes the weakest user first, and then subtracts its component from the received signal. In the second iteration, the user decodes the second weakest user using the remaining signal. The cancellation iterates until each user gets its own signal.
\end{itemize}

First, we observe the numerical results of a simple setup with $2$ vehicle surfaces and $2$ users, as shown in \figurename \ref{Fig_results1}.  In our simulations, the number of iterations for alternating optimization is set to three, which is sufficient for the convergence of optimization. The $95\%$-likely spectral efficiency, which is usually applied to measure the performance of cell-edge users, and the $50\%$-likely or median spectral efficiency of different configurations are listed in Table \ref{table_IRSdatarates}. As we expected, NOMA is superior to the conventional TDMA due to advanced signal processing (i.e., superposition coding and SIC). It is observed that the use of vehicle surfaces brings substantial performance improvement for such a moving vehicle network.  Even though random phase shifts are applied, both the $95\%$-likely and $50\%$-likely rates increase approximately \SI{7}{\bps\per\hertz^{}} compared with the absence of IRVS. If the phase shifts are smartly adjusted, even 1 phase-control bit can bring another growth of around \SI{7}{\bps\per\hertz^{}}.
Second, we also illustrate the performance evaluation of these schemes in the case of $20$ users with the aid of $5$ surfaces, as shown in \figurename \ref{Fig_results2}. We observe that the distribution of achievable sum rate becomes more concentrated, and the superiority of NOMA enlarges, with the increasing number of active users. In this case, the use of vehicle surfaces still bring an obvious performance improvement over NOMA and TDMA, especially for the $95\%$-likely spectral efficiency. The IRVS with 2 phase-control bits results in four-fold and five-fold rate increase of $95\%$-likely and $95\%$-likely spectral efficiencies, respectively. In a nutshell, the numerical results justify the effectiveness and superiority of applying IRVS in moving vehicular networks.

\section{Conclusions}
This paper focused on applying the technology of intelligent reflection surface to boost the capacity of moving vehicular networks in military and emergency communications. We presented a novel paradigm coined Intelligent Reflecting Vehicle Surface or IRVS, which integrates a massive number of reflection elements on the surfaces of vehicles. Moreover, an alternative optimization method was proposed to optimize jointly active beamforming at the base station and passive reflection at the vehicle surfaces. Performance evaluation in terms of sum spectral efficiency under continuous, discrete, and random phase shifts was conducted, in comparison with the performance of the conventional TDMA and NOMA schemes in the absence of IRVS. Numerical results revealed that the introduction of IRVS can substantially improve the capacity of a moving vehicular network, even if the IRVS operates in a blind mode with random phase shifts.




%

\bibliographystyle{IEEEtran}
\bibliography{IEEEabrv,Ref_Milcom}

\end{document}